\documentclass[twocolumn]{autart}    

\pdfoutput=1

\usepackage{cite}

\usepackage{graphicx}
\usepackage{amsmath}
\usepackage{cite}
\usepackage{mathtools, amssymb}
\usepackage{varioref}
\labelformat{equation}{(#1)}
\usepackage{amsfonts}

\usepackage{color}
\usepackage{enumerate}

\usepackage{mysymbol}

\usepackage[dvipsnames]{xcolor}
\usepackage{tikz,pgfplots}
\usepackage{color} 
\usetikzlibrary{matrix} 
\usetikzlibrary{arrows} 
\usetikzlibrary{calc} 
\usetikzlibrary{shapes}
\pgfplotsset{compat=1.7}

\usepackage{standalone}
\usetikzlibrary{arrows}
\usetikzlibrary{positioning}
\usetikzlibrary{backgrounds, fit}

\theoremstyle{plain}
\newtheorem{lemma}{Lemma}
\newtheorem{proposition}{Proposition}

\theoremstyle{definition}

\newtheorem{remark}{Remark}

\newcommand{\qedsymbol}{\hspace*{\fill}~$\blacksquare$}

\renewcommand{\epsilon}{\varepsilon}

\newcommand{\figref}[1]{Fig.~\ref{fig:#1}}

\usepackage{algorithm}
\usepackage{algpseudocode}
\usepackage{algorithmicx}

\usepackage{epstopdf}

\begin{document}
	
	\begin{frontmatter}
		
\title{
Statistical Learning for Analysis of Networked Control \\Systems over Unknown Channels
}

\author{Konstantinos Gatsis}\ead{kgatsis@seas.upenn.edu}, 
\author{George J. Pappas}\ead{pappasg@seas.upenn.edu}
\address{Department of Electrical and Systems Engineering, University of Pennsylvania, 200 South 33rd Street, Philadelphia, PA 19104}  
\thanks{
	This research is partially supported by NSF CPS-1837253, by the Intel Science and Technology Center for Wireless Autonomous Systems (ISTC-WAS)
and by ARL DCIST CRA W911NF-17-2-0181. The authors are with the Department of Electrical and Systems Engineering, University of Pennsylvania, 200 South 33rd Street, Philadelphia, PA 19104. Email: \{kgatsis, pappasg\}@seas.upenn.edu.}%


\begin{abstract}
Recent control trends are increasingly relying on communication networks and wireless channels to close the loop for Internet-of-Things applications. Traditionally these approaches are model-based, i.e., assuming a network or channel model they are focused on stability analysis and appropriate controller designs. However the availability of such wireless channel modeling is fundamentally challenging in practice as channels are typically unknown a priori and only available through data samples. In this work we aim to develop algorithms that rely on channel sample data to determine the stability and performance of networked control tasks. In this regard our work is the first to characterize the amount of channel modeling that is required to answer such a question. Specifically we examine how many channel data samples are required in order to answer with high confidence whether a given networked control system is stable or not. This analysis is based on the notion of sample complexity from the learning literature and is facilitated by concentration inequalities. Moreover we establish a direct relation between the sample complexity and the networked system stability margin, i.e., the underlying packet success rate of the channel and the spectral radius of the dynamics of the control system. This illustrates that it becomes impractical to verify stability under a large range of plant and channel configurations. We validate our theoretical results in numerical simulations.
\end{abstract}

\end{frontmatter}

\section{Introduction}

Wireless communication is increasingly used in autonomous applications to connect devices in industrial control environments, teams of robotic vehicles, and the Internet-of-Things. To guarantee safety and control performance it is customary to include a model of the wireless channel, for example an i.i.d. or Markov link quality, alongside the model of the physical system to be controlled. In such modeled-based approaches one can characterize, for example, that it is impossible to estimate or stabilize an unstable plant if its growth rate is larger than the rate at which the link drops packets~\cite{Sinopoli_intermittent, Schenato_foundations, hespanha2007survey}, or below a certain channel capacity~\cite{Tatikonda, Sahai_anytime}. Models also facilitate the allocation of communication resources to optimize control performance in, e.g., power allocation and scheduling over fading channels~\cite{GatsisEtal14, Quevedo_Kalman, GatsisEtal15}, or in event-triggered control~\cite{mazo2011decentralized, araujo2014system, mamduhi2014event}. 

In practice wireless autonomous systems operate under unpredictable channel conditions following unknown distributions, which are more often observable via a finite amount of collected channel sample measurements~\cite{halperin2010predictable,rappaport2015wideband}. 
The purpose of this work is the analysis of networked control systems when only channel sample data are available instead of channel models.
We use the data to learn whether a given networked control system is stable, and 
we also characterize how the learning procedure depends on the amount of channel samples and the control system parameters.
%
%
To the best of our knowledge, our paper is the first to consider the sample complexity analysis of data-based algorithms for networked control, in contrast to the extensive literature on model-based approaches mentioned above.

Learning methods have been used in control problems most commonly within the reinforcement learning and approximate dynamic programming literature~\cite{sutton1998reinforcement, bertsekas}, where the goal is to learn to control an unknown dynamical system from data. One specific approach within this framework is based on learning the system dynamics model first~\cite{kumar2015stochastic}. This is used for example in analyzing the sample complexity of the classic multi-armed bandit problem~\cite{lai1985asymptotically,auer2002finite} and more recently in the quadratic control of unknown linear systems~\cite{abbasi2011regret, dean2017sample, ouyang2017learning, tu2018least}. Other approaches focus on learning the controller directly as in policy gradient methods~\cite{sutton1998reinforcement}.  In contrast our work is focused on collecting data and learning unknown channel models instead of system dynamics. { In the context of networked control systems very recent works from the last two years are proposing data-based approaches including deep learning for allocating resources and scheduling \cite{demirel2018deepcas,redder2019deep,leong2018deep,wu2018learning, eisen2019control} as well as for controller design \cite{schuurmans2019safe,trimpe}.} A related but broader topic is also the identification of switched systems \cite{ozay2011sparsification} where the matrix dynamics are also unknown. To the best of our knowledge our work is the first to examine the sample complexity of networked control systems analysis and the amount of channel modeling needed for such problems. We also point out that an alternative approach is to bypass building channel models altogether and learn solutions directly as in our previous work on power allocation in~\cite{EisenEtal19} and multiple-access in \cite{GatsisEtal15, GatsisEtal18}.

Specifically we consider the stability of a linear dynamical system over a Bernoulli packet-dropping channel with an unknown success rate (Section~\ref{sec:problem}). Using channel sample data, i.e., a number of packet successes and failures, we develop an algorithm to learn whether the networked control system is stable or not (Section~\ref{sec:results}). To do this we utilize confidence bounds obtained by concentration inequalities, more specifically, Hoeffding's inequality. As our algorithm depends on random channel samples there is always a probability of error, i.e., the algorithm returns that the system is stable while the true system is not. Our algorithm is guaranteed by design to have low probability of such errors \textit{irrespective of the number of data points} provided. On the other hand we characterize the statistical correctness of the algorithm (Theorem~\ref{thm:error_bound}), i.e., the probability of correctly learning the stability of the networked control system. We further analyze the amount of channel sample data needed to correctly learn the system stability with a desired confidence level (Corollary~\ref{cor:sample_stability}).

Our most significant finding is that the sample complexity adversely depends on the system stability margin, i.e., the underlying packet success rate of the channel and the spectral radius of the dynamics of the control system. In other words, a {significantly larger} number of samples are needed to learn whether the system is stable if the networked control system over the channel is {closer} to instability. This means that it becomes impractical to verify stability under a large range of plant and channel configurations, but the derived sample complexity can be useful as follows. It describes the amount of channel samples required if we are willing to verify stability with high confidence up to a certain system stability margin. We show that a tighter sample complexity analysis can be performed for very reliable or very unreliable links (Section~\ref{sec:tighter}). We show that the nature of the results remains the same when instead of analyzing stability we are interested in testing whether the (quadratic) cost of the networked system is below some test value (Section~\ref{sec:control_performance}). We validate our theoretical analysis in numerical simulations (Section~\ref{sec:numerical}). A preliminary version of the results appeared in \cite{CDC18_GatsisPappas}.

\section{Model-based Networked Control}\label{sec:problem}

\begin{figure}[!t]
	\centering
	\includegraphics[width = \columnwidth]{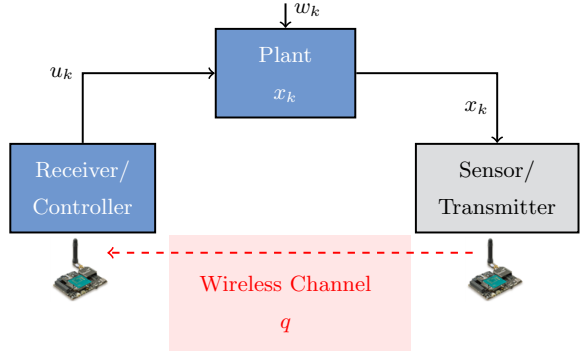}
	\vspace{0pt}
	\caption{
		Wireless Control System. A sensor measures the state of a plant perturbed by a random disturbance. The sensor transmits the measured information over a packet-dropping wireless channel to a receiver/controller providing control inputs.
	}
	\vspace{0pt}
	\label{fig:wireless_control}
\end{figure}

We consider the evolution of a dynamical system over a packet dropping channel. This is a standard model for remote estimation or control over a network or a wireless channel, for example when a sensor measures the state of the plant and transmits it to a receiver -- see~\figref{wireless_control} and~\cite{Sinopoli_intermittent, Schenato_foundations, hespanha2007survey} for related examples. 
Our goal is to analyze the stability properties and the control performance of the system, hence we assume the dynamics for the system are fixed, for example a controller has been already designed. We assume that the evolution of the system depends on whether a transmission occurs at time $k$ or not, indicated with variables $\gamma_{k} \in \{0,1\}$. 
We suppose the system evolution is described by a switched linear time invariant model of the form
\begin{equation}\label{eq:system}
	x_{k+1} = \left\{ \begin{array}{ll}
	A \, x_{k} + w_{k}, &\text{ if } \gamma_{k}=0 \\
	 w_{k}, &\text{ if } \gamma_{k}=1
	\end{array}\right. .
\end{equation}
Here $x_{k} \in \reals^{n}$ denotes the state of the overall control system at each time $k$. At a successful transmission the system dynamics are reset to zero, and otherwise when the transmission fails the dynamics are in open loop described by $A \in \reals^{n \times n}$. 
The open loop matrix $A $ may be unstable, i.e., the eigenvalue with the largest modulus may be larger than unity, $\rho(A) = \max_{i=1, \ldots, n} \, |\lambda_i(A)| >1 $. This case motivates the stability analysis. The additive terms $w_{k}, k \geq 0$ model an independent identically distributed (i.i.d.) noise process across time according to some known probability distribution 
 with mean zero and positive definite covariance $W$. 

We are interested in the performance of the dynamical system. We will employ the usual quadratic system state cost at each time step $k$ as $x_k^T \, Q \, x_k $
where $Q$ is a positive definite matrix. Intuitively there is a higher penalty when the state of the system is away from the origin. 

The cost over time depends on whether the transmissions are successful or not over time. In this paper we make the assumption that $\gamma_k$ is an independent Bernoulli random variables with a constant success probability $q$, and they are also independent from the system state $x_k$ and noise $w_k$. This i.i.d. assumption is very crucial for our results as we discuss in Section~\ref{sec:discussion}.

Given the model of the transmission success we can describe the effect on the performance of the control system with the long turn average quadratic cost
\begin{align}\label{eq:control_cost}
	J(q) = \limsup_{N \rightarrow \infty} \frac{1}{N} \sum_{k=0}^{N-1} 
	\mathbb{E} [ x_k^T \, Q \, x_k ] .
\end{align}
The expectation at the right hand side accounts for the randomness introduced by the system disturbance and the channel. We choose to denote this cost as a function of the success rate of the channel $q$.

When the channel success rate $q$ is known, we have the following fundamental result.

\begin{proposition}\label{prop:model_based}
Consider the switched linear system \ref{eq:system} over an i.i.d. Bernoulli binary channel with a known success probability $q \in [0,1]$. Then:\\
\begin{enumerate}
\item The system is stable, i.e., $\sup_k \mathbb{E}x_k x_k^T <\infty$ if and only if 
\begin{equation}\label{eq:stability_condition}
	q> 1 - \frac{1}{\rho(A)^2}
\end{equation}
\item Moreover, if the condition \ref{eq:stability_condition} holds, the quadratic control cost \ref{eq:control_cost} is a 
{strictly decreasing} function of the success rate $q$ given by
\begin{equation}\label{eq:cost_function_of_q}
	J(q) = \text{Tr}(P W)
\end{equation}
where $P$ is the unique positive definite solution of the (Lyapunov) matrix equation
\begin{equation}\label{eq:Lyapunov_equation}
	P = Q + (1-q) A^T P A.
\end{equation}
\end{enumerate}
\end{proposition}

\begin{pf*}{Proof.}
	The stability condition \ref{eq:stability_condition} as well as the expression for the control cost in \ref{eq:cost_function_of_q}-\ref{eq:Lyapunov_equation} follow from the random jump linear system theory~\cite{Costa_Fragoso_book}. 
	
	The fact that the control cost \ref{eq:control_cost} is a decreasing function can be shown as follows. It is straightforward to show that the solution to the Lyapunov equation \ref{eq:Lyapunov_equation} can be written as 
\begin{equation}
	P = \sum_{i=0}^\infty \; (1-q)^i \, (A^i)^T Q A^i,
\end{equation}
where the sum converges due to \ref{eq:stability_condition}. Plugging in this expression in \ref{eq:cost_function_of_q} yields the expression for the control cost as 
\begin{equation}
	J(q) = \sum_{i=0}^\infty \; (1-q)^i \, \text{Tr}( (A^i)^T Q A^i W) .
\end{equation}
Taking the derivative with respect to $q$ verifies that the function is {non-increasing} because all terms $\text{Tr}( (A^i)^T Q A^i W) $ are non-negative { and is strictly decreasing because $\text{Tr}( Q W) =\text{Tr}( W^{1/2}Q W^{1/2}) >0 $ as both $Q, W$ are positive definite matrices.}. 
\qedsymbol
\end{pf*}

The above result is a fundamental limit in the sense that it characterizes the absolute minimum channel success rate required for stability as a function of the eigenvalues of the dynamics $A$. The proposition also gives an expression for the control performance as a function of the system and channel parameters. 

However in practice the channel success rate $q$ is unknown. Instead we may have access to channel sample data. The problem we would like to answer is twofold:
\begin{itemize} 
\item Data-driven Stability Analysis: how to check whether the system is stable or not using the channel sample data? 
\item Sample Complexity: what is the confidence of the method and how does it scale with the amount of channel samples and the control system and channel parameters?
\end{itemize}


\section{Sample-based Stability Analysis of Networked Control}\label{sec:results}

Suppose that instead of knowing the packet success rate $q$ of the channel we have available $N$ samples denoted by $\gamma_{k}, k=0, \ldots, N-1$ drawn independently from the Bernoulli distribution with success $q$. In practice this data is easy to obtain, it suffices to send $N$ packets and record whether they are received or not. Given this data the most natural approximation of the true success probability $q$ is the sample average
\begin{equation}
	\hat{q}_N = \frac{1}{N} \sum_{k=0}^{N-1} \gamma_k .
\end{equation}
Indeed this approximation is in some sense optimal\footnote{It is easy to show that the sample average is the maximum likelihood estimate of the underlying channel success rate $q$. The likelihood of some success rate $q \in [0,1]$ give the data $\gamma_{k}, k=0, \ldots, N-1$ is given by the probability that the data would be obtained under this success rate $q$. That is given by $\mathbb{P}(\gamma_{k}, k=0, \ldots, N-1 \given q) = q^{\sum_{k} \gamma_k} \, (1-q)^{(N - \sum_k \gamma_k)}$. Maximizing the likelihood (or its logarithm) readily gives the sample mean as the most likely packet success.}.

In the case of unlimited data samples the sample average converges almost surely to the true underlying packet success rate by the Strong Law of Large Numbers~\cite[Ch.2]{Durrett_probability}. 
Hence in the face of unlimited data, learning the stability of the control system, i.e., checking whether \ref{eq:stability_condition} holds, would be easy. However this is an asymptotic analysis. In practice only finite amount of data will be available and this motivates us to investigate a finite sample analysis.

For a finite number of samples we argue that instead of point estimates of the channel success rate, confidence intervals are more useful. We further characterize the sample complexity, i.e., how many channel samples are needed to verify the stability of the system.

Our approach is based on confidence intervals constructed by the channel sample data using concentration inequalities. In particular we may  employ Hoeffding's inequality.

\begin{lemma}\label{lemma:Hoeffding}[Hoeffding's inequality, Th. 2.8~\cite{boucheron2013concentration}] Consider a sequence $\{\gamma_k, k=0, \ldots, N-1\}$ of i.i.d. random variables taking values in $[0,1]$ with mean $q$. Let $\hat{q}_N = \frac{1}{N}\sum_{k=0}^{N-1} \gamma_k$ be the sample average. Then for any $\epsilon >0$ we have that 
\begin{align}
		&\mathbb{P}( \hat{q}_N \geq q +  \epsilon ) \leq \exp\{-2 N \epsilon^2\},\\
				&\mathbb{P}( \hat{q}_N \leq q -  \epsilon ) \leq \exp\{-2 N \epsilon^2\},
\end{align}  
where the probability is with respect to the random sequence $\{\gamma_k, k=0, \ldots, N-1\}$.
\end{lemma}

The result essentially states that there is a low probability that the sample average deviates much from the true packet success rate and further provides an explicit bound on this probability. {Note that the result holds regardless of the distribution as long as it has a bounded support. In the particular case we consider the result can be strengthened as the distribution of the sum of i.i.d. Bernoulli random variables $\sum_k \gamma_k$ is known to be binomial so more precise approaches may be employed -- see Section \ref{sec:binomial}. Here we opt for the  bound above for simplicity. In numerical simulations in Section~\ref{sec:numerical} we will also examine its conservativeness.} 

The important aspect of the above result is that it reveals a closed form dependency between the number of samples $N$, the deviation amount $\epsilon$, and the confidence level which is the bound at the right side of the inequality. There is a useful direct consequence of this inequality. Given a desired high confidence level $1 - \delta$ where $\delta$ is a small positive value, for example of the order of $10^{-3}$, and after collecting $N$ samples, we may derive an interval where the true underlying mean lies, that is, the channel success rate in our problem, as follows.

\begin{lemma}\label{lemma:Hoeffding_2}
Consider a sequence $\{\gamma_k, k=0, \ldots, N-1\}$ of i.i.d. random variables taking values in $[0,1]$ with mean $q$. Let $\hat{q}_N = \frac{1}{N}\sum_{k=0}^{N-1} \gamma_k$ be the sample average. Then for any $\delta \in (0,1)$ it holds that 
\begin{align}
&\mathbb{P}\left(q \geq 
 \hat{q}_N - \sqrt{\frac{\log(1/\delta)}{2 N}}
\right) \geq 1 - \delta,
\end{align}  
%
where the probability is with respect to the random sequence $\{\gamma_k, k=0, \ldots, N-1\}$.
\end{lemma}

In this lemma the quantity $\hat{q}_N - \sqrt{\frac{\log(1/\delta)}{2 N}}$ is a sample-based high-confidence lower bound on the true packet success rate of the channel. We also note the dependency on the amount of samples. Doubling the amount of collected channel samples from $N$ to $2N$ only slightly tightens the lower bound (by a factor of $1/\sqrt{2}$).

\subsection{Stability Analysis Using Channel Samples}

Let us now return to the main question of this paper. Given some channel data we would like to verify whether the system is stable, that is, whether the inequality \ref{eq:stability_condition} holds. We propose to utilize Hoeffding's inequality. We can construct an interval where the channel success rate lies with a desired high confidence using Lemma~\ref{lemma:Hoeffding_2}. Then we can check whether stability holds for all such high-confidence channel conditions. In particular it suffices to check whether stability holds for the lower end of this interval. A symmetric argument can verify instability of the system. This data-based procedure is summarized in Algorithm 1.


\begin{algorithm}[!t]
\caption{Stability analysis using channel samples}\label{alg:stability}
\begin{algorithmic}[1]

\Require Dynamics $A$, Confidence level $\delta$, Number of samples $N$, Channel samples $\gamma_0, \ldots, \gamma_{N-1} \in \{0,1\}^N$
\State Compute the sample average
\begin{equation}
	\hat{q}_N = \frac{1}{N} \sum_{k=0}^{N-1} \gamma_k
\end{equation}
\State Compute the high confidence lower and upper bounds
\begin{align}
	q_{\min} &= \hat{q}_N - \sqrt{\frac{\log(1/\delta)}{2 N}}\\
	q_{\max}& = \hat{q}_N + \sqrt{\frac{\log(1/\delta)}{2 N}}
\end{align}
\If{ $1 - \frac{1}{\rho(A)^2} < q_{\min}$ } 
\State\Return 'Stable' 
\Else
\If{ $1 - \frac{1}{\rho(A)^2} > q_{\max}$ } 
\State\Return 'Unstable' 
\Else {} 
	\State\Return 'Undetermined' 
\EndIf
\EndIf

\end{algorithmic}
\end{algorithm}

{First we note that it is possible that the algorithm returns a wrong answer, i.e., return 'Stable' in cases where the system is unstable and vice versa. Intuitively this can occur if the sample mean of the data is sufficiently different from the true mean. However \textit{by design} we can guarantee that wrong answers are very unlikely (happen with probability less than $\delta$) and also \textit{irrespective of the number of data points} $N$ provided, because of the use of the confidence intervals.}
On the other hand the algorithm may not be able to correctly determine stability or instability for small amount of data, for example, return 'Undetermined' in cases where the system is actually stable. The remedy is to draw more samples and check stability again. By the Strong Law of Large Numbers the sample mean and the upper and lower high-confidence bounds will all converge to the true value $q$ and stability will be asymptotically determined correctly.

Instead of asymptotic performance, our main result is to analyze the average performance of Algorithm 1 using a finite number of samples. By average performance, we mean how often would the algorithm return the correct or the wrong answer if we were to run it multiple times over independent data samples. We formalize this next.

Let $A_N$ and $W_N$ denote the events that algorithm provides a correct answer and wrong answer respectively. That is, if the system is stable, i.e., condition $q > 1 - \frac{1}{\rho(A)^2} $ holds, then define te event
\begin{equation}\label{eq:correct_stability}
	A_N = \Big\{ \{\gamma_0, \ldots, \gamma_{N-1} \} \in \{0,1\}^N :\text{ Alg. 1 returns 'stable'} \Big\}
\end{equation}
Similarly define the event
\begin{equation}\label{eq:wrong_stability}
W_N = \Big\{ \{\gamma_0, \ldots, \gamma_{N-1} \} \in \{0,1\}^N :\text{Alg. 1 returns 'unstable'} \Big\}
\end{equation}
The complement of $A_N$ and $W_N$ is the event that the algorithm returns the answer 'Undetermined'.

Alternatively if the system is strictly unstable, i.e., condition $q < 1 - \frac{1}{\rho(A)^2} $ holds, then we have to define the events as
\begin{equation}\label{eq:correct_instability}
A_N = \Big\{\{\gamma_0, \ldots, \gamma_{N-1} \} \in \{0,1\}^N  :\text{Alg. 1 returns 'unstable'} \Big\}
\end{equation}
and
\begin{equation}\label{eq:wrong_instability}
W_N = \Big\{ \{\gamma_0, \ldots, \gamma_{N-1} \} \in \{0,1\}^N :\text{ Alg. 1 returns 'stable'} \Big\}
\end{equation}
For convenience and because of symmetry we use the same notation for the events in the two cases above, even though these are distinct  events. Alternatively the two cases could be examined separately. We can state then the following main result.

\begin{thm}[Sample-based Stability Analysis] \label{thm:error_bound}
Consider the switched linear system \ref{eq:system} over an i.i.d. Bernoulli binary channel with an unknown success probability $q \in [0,1]$ and assume $q \neq 1 - \frac{1}{\rho(A)^2}$. Consider the stability analysis procedure developed in Algorithm \ref{alg:stability} using $N$ i.i.d. channel samples drawn with success rate $q$. Let $A_N$ denote the event that the algorithm provides the correct answer as defined in \ref{eq:correct_stability} or \ref{eq:correct_instability}, and let $W_N$ denote the event that the algorithm provides the wrong answer as defined in \ref{eq:wrong_stability} or \ref{eq:wrong_instability}. Then for any $N \geq 1$ it holds that\footnote{here $[\;]_+$ denotes the projection to the positives}
	\begin{align}\label{eq:error_bound}
	&\mathbb{P}(A_N) 
	\geq \nonumber\\
	&1 - \exp\left\{-2 N \left[\left| q- 1 + \frac{1}{\rho(A)^2} \right| - \sqrt{\frac{\log(1/\delta)}{2 N}} \right]_+^2\right\}
	\end{align}
	and
	\begin{align}\label{eq:wrong_error_bound2}
	&\mathbb{P}(W_N) 
	\leq \delta
	\end{align}
	where the probability is with respect to the random channel samples.
\end{thm}

\begin{pf*}{Proof.}
Suppose the system is stable, i.e., according to Proposition~\ref{prop:model_based} the packet success probability satisfies
\begin{equation}\label{eq:stability_condition_here}
	q > 1 - \frac{1}{\rho(A)^2} .
\end{equation}
The event $A_N$ that Algorithm~\ref{alg:stability} returns the correct result is defined by \ref{eq:correct_stability}, and in this case corresponds to the case  
\begin{align}\label{eq:correct_stability_1}
	A_N = \{ &\{\gamma_0, \ldots, \gamma_{N-1} \} \in \{0,1\}^N \text{ s.t. }  \\
	&\hat{q}_N - \sqrt{\frac{\log(1/\delta)}{2 N}} > 1 - \frac{1}{\rho(A)^2}
 \}.
\end{align}
As a result we have that 
\begin{align}\label{eq:correct_stability_2}
	\mathbb{P}(A_N) = 1 - \mathbb{P}\left[ \hat{q}_N - \sqrt{\frac{\log(1/\delta)}{2 N}} 
	\leq 1 - \frac{1}{\rho(A)^2}
 \right].
\end{align}
Adding and subtracting $q$ at the right hand side we have that
\begin{align}\label{eq:correct_stability_3}
	\mathbb{P}(A_N) = 1 - \mathbb{P}\left[ \hat{q}_N 
	\leq q - \left(q - 1 + \frac{1}{\rho(A)^2} - \sqrt{\frac{\log(1/\delta)}{2 N}}\right)
 \right]
\end{align}
The term in the parenthesis can be in general both negative or positive, hence we consider two cases.

\noindent Case I: $q - 1 + \frac{1}{\rho(A)^2} - \sqrt{\frac{\log(1/\delta)}{2 N}} >0$. This is the case where the term in the parenthesis in \ref{eq:correct_stability_3} is positive and we can directly apply Hoeffding's inequality (Lemma~\ref{lemma:Hoeffding}) to get the desired bound \ref{eq:error_bound}.

\noindent Case II: $q - 1 + \frac{1}{\rho(A)^2} - \sqrt{\frac{\log(1/\delta)}{2 N}} \leq 0$. In this case the bound in \ref{eq:error_bound} becomes
\begin{equation}
	\mathbb{P}(A_N) \geq 1 - \exp\{ - 2 N \, 0\} = 0,
\end{equation}
which trivially holds.

Using similar arguments we can show that the algorithm provides the wrong answer with probability bounded by
\begin{align}\label{eq:wrong_error_bound}
&\mathbb{P}(W_N) 
\leq \exp\left\{-2 N \left[\left| q- 1 + \frac{1}{\rho(A)^2} \right| + \sqrt{\frac{\log(1/\delta)}{2 N}} \right]^2\right\}
\end{align}
In particular since the term in the absolute value is non-negative, the bound \ref{eq:wrong_error_bound2} can be obtained by bounding this absolute value below by zero.

A symmetric argument verifies the bound \ref{eq:error_bound} when the system is strictly unstable, i.e., when the packet success probability satisfies $q < 1 - \frac{1}{\rho(A)^2}$.
\qedsymbol\end{pf*}

Some remarks are in order. First, the probability that the algorithm returns the correct answer grows to one as the number of samples grows to infinity. This is expected from the Law of Large Numbers as already mentioned. But for finite number of samples there will be a probability the algorithm may not be confident enough to correctly identify stability. More importantly, this probability depends on how far the system is from stability, i.e., the term $\left| q- 1 + \frac{1}{\rho(A)^2} \right| $. The largest the stability margin the easier it is to get the correct result and vice versa. {As already mentioned, by design the algorithm is very unlikely to provide a wrong answer as verified by \ref{eq:wrong_error_bound2}. }

The theorem also assumes that $q \neq 1 - \frac{1}{\rho(A)^2}$, i.e., that there is a non-zero stability margin in view of Proposition~\ref{prop:model_based}. Technically the reason is that in that case while $\hat{q}_N$ converges to $q$ it may take values both above and below the limit $q$ and hence the algorithm may oscillate between the answers 'Stable' and 'Unstable' as the number of samples increases. The assumption is practically not restrictive, it only excludes a measure zero case for the channel packet success rate.

The above result is important as it allows us to characterize the finite sample complexity of the algorithm, i.e., answer how many channel sample data we need in order to correctly determine stability with a desired high confidence given the system parameters.

\begin{cor}[Sample Complexity]\label{cor:sample_stability} 
Consider the switched linear system \ref{eq:system} over an i.i.d. Bernoulli binary channel with an unknown success probability $q \in [0,1]$ and assume $q \neq 1 - \frac{1}{\rho(A)^2}$. Consider the stability analysis procedure developed in Algorithm \ref{alg:stability} using $N$ i.i.d. channel samples drawn with success rate $q$ and some parameter $\delta \in (0,1)$. If the number of samples is 
\begin{equation}\label{eq:sample_complexity}
N \geq \frac{2 \log(1/\delta)}{ (q- 1 + \frac{1}{\rho(A)^2})^2 } ,
\end{equation}
then the procedure correctly determines the stability or instability of the system with probability $(1-\delta)$.
\end{cor}

\begin{pf*}{Proof.}
Let the number of samples satisfy \ref{eq:sample_complexity}. Then it holds that
\begin{equation}
	\left| q- 1 + \frac{1}{\rho(A)^2} \right| \geq \sqrt{\frac{2 \log(1/\delta)}{ N}}
	= 2 \sqrt{\frac{ \log(1/\delta)}{ 2 N}},
\end{equation}
which is equivalent to
\begin{equation}
	\left[\left| q- 1 + \frac{1}{\rho(A)^2} \right| - \sqrt{\frac{\log(1/\delta)}{2 N}} \right]_+
	\geq \sqrt{\frac{ \log(1/\delta)}{ 2 N}},
\end{equation}
or
\begin{equation}
	-2 N \left[\left| q- 1 + \frac{1}{\rho(A)^2} \right| - \sqrt{\frac{\log(1/\delta)}{2 N}} \right]_+^2
	\leq - \log(1/\delta).
\end{equation}
Hence by the Theorem~\ref{thm:error_bound} we conclude that the Algorithm~\ref{alg:stability} returns the correct answer with probability at least $1-\delta$.
\qedsymbol\end{pf*}

The number of samples required depends on the true channel success rate $q$ which is unknown so it is not directly useful but provides intuition. We observe from this result that the sample complexity scales well with the desired confidence level. An order of magnitude improvement in confidence can be guaranteed with just doubling the amount of data samples. On the other hand, the sample complexity does not scale well with the system stability margin. Reducing the stability margin $|q- 1 + \frac{1}{\rho(A)^2}|$ by a factor of $\beta$ requires $\beta^2$ more channel samples. 

\begin{figure}[!t]
	\centering
	\includegraphics[width=1.0\columnwidth]{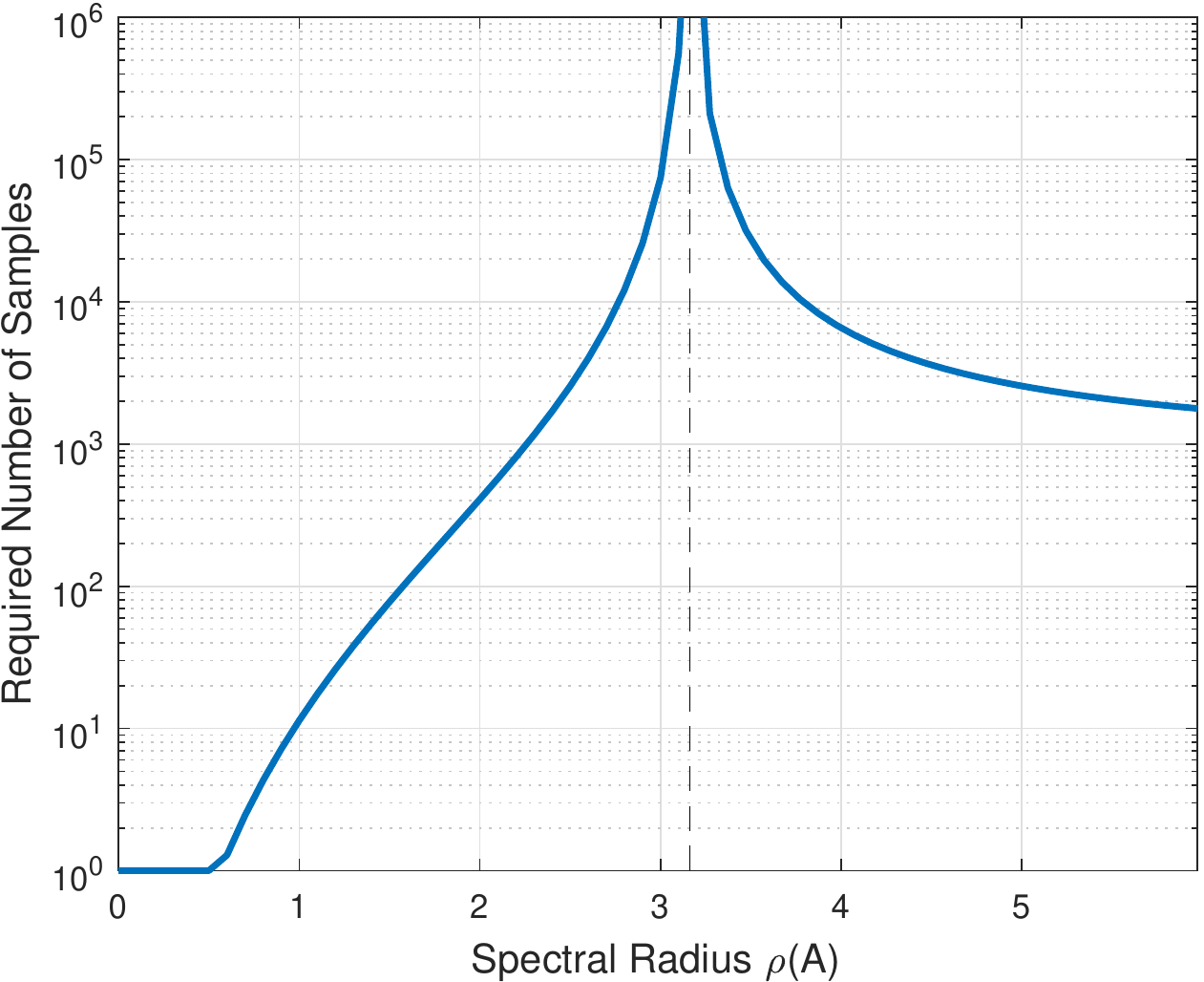}
	\vspace{-0pt}
	\caption{Sample complexity according to Corollary~\ref{cor:sample_stability}. For a channel with fixed success rate $q = 0.9$ and a desired confidence $1 - \delta = 0.99$ we plot the theoretically required number of channel samples $N$ as a function of the system spectral radius. The required number of samples grows unbounded as the stability radius reaches the critical point of instability, i.e., when $q = 1 - \frac{1}{\rho(A)^2}$, denoted by the dotted line.}
	\label{fig:N_required_theory}
	\vspace{-0pt}
\end{figure}

\begin{figure}[!t]
	\centering
	\includegraphics[width=1.0\columnwidth]{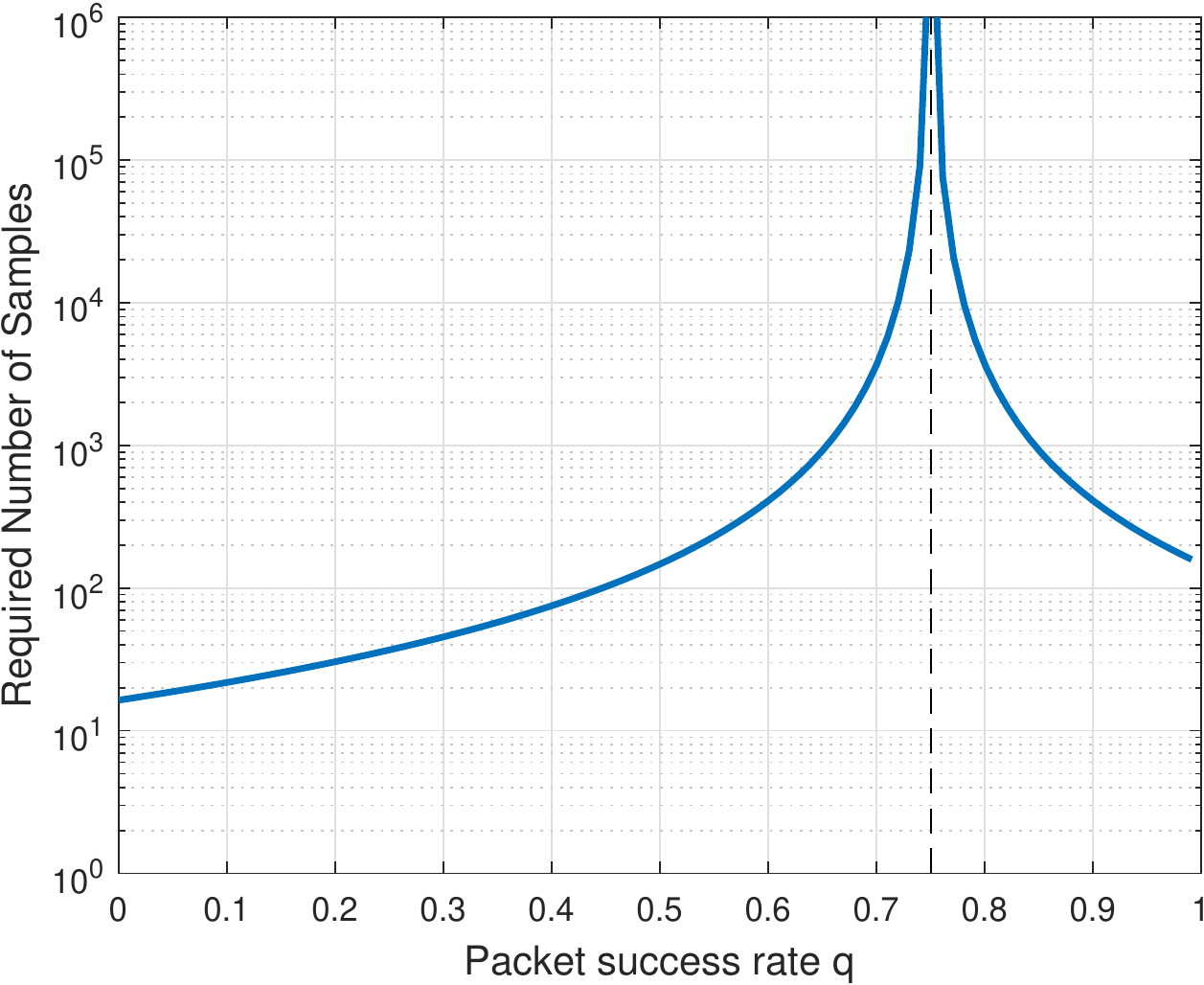}
	\vspace{-0pt}
	\caption{Sample complexity according to Corollary~\ref{cor:sample_stability}. For a system with fixed spectral radius $\rho(A) = 2$ and a desired confidence $1 - \delta = 0.99$ we plot the theoretically required number of channel samples $N$ as a function of the packet success rate $q$ of the channel. The required number of samples grows unbounded as the packet success reaches the critical point of instability, i.e., when $q = 1 - \frac{1}{\rho(A)^2}$, denoted by the dotted line.}
	\label{fig:N_required_theory_q}
	\vspace{-0pt}
\end{figure}

A plot for this sample relation \ref{eq:sample_complexity} is given in Fig.~\ref{fig:N_required_theory} as a function of the system spectral radius $\rho(A)$ and for a fixed channel success rate $q$. For very slow dynamics the algorithm provides a correct answer trivially even with a single sample. More importantly the required number of samples grows unbounded as the stability radius reaches the critical point of instability, i.e., when $q = 1 - \frac{1}{\rho(A)^2}$. Moreover, if the system dynamics are very fast, then the algorithm can correctly verify that the system is unstable with fewer channel samples. Similarly in Fig.~\ref{fig:N_required_theory_q} we plot the theoretically required number of samples \ref{eq:sample_complexity} for a fixed system $\rho(A)$ as a function of the underlying packet success rate $q$ of the channel. We observe similar results, i.e., that the number of samples needs to grow unbounded close to the critical point where the stability margin vanishes.

These observations mean that it becomes impractical to verify stability under all plant and channel configurations, but the above sample complexity can be useful as follows. It describes the amount of channel samples required if we are willing to verify stability with high confidence up to a certain system stability margin. 

{\begin{remark} The bound in \ref{eq:sample_complexity} can be thought as asymptotically tight in the following sense. For very large values $N$ by the Central Limit Theorem the variables ${q}_{\min}$ and ${q}_{\max}$ are both distributed as normal distributions centered at the true mean $q$ with a variance $\sigma^2 = q(1-q)\sqrt{N}$. The stability test that we conduct checks whether $\hat{q}_{\max}< 1 - 1/\rho(A)^2$ or  $\hat{q}_{\min}> 1 - 1/\rho(A)^2$. Hence it gives the correct answer with probability $\mathbb{P}(A_N) \approx 1 - \mathbb{P}( \mathcal{N}_{0, \sigma} \geq |q- 1 + \frac{1}{\rho(A)^2}| ) \approx 1 - \exp( - |q- 1 + \frac{1}{\rho(A)^2}|^2/ (q(1-q)\sqrt{N} )^2$. From this we conclude that the probability of correct answer increases to unity exponentially with a rate $|q- 1 + \frac{1}{\rho(A)^2}|^2/ N$.
\end{remark}}

\subsection{Sharper sample complexity results for low-variance channels}\label{sec:tighter}

The previous analysis uses Hoeffding's inequality to obtain a sample complexity result for any potential channel success rate $q\in [0,1]$ we may encounter. In this section we will show that using sharper concentration inequality bounds we can obtain more advantageous results in the case that we are dealing with very reliable or very unreliable links ($q\approx 0$ or $q \approx 1$). In particular in these cases we will show that fewer samples are needed to obtain high performance in the data-driven stability problem. This may be practically important in cases where one a priori knows that we are dealing with very reliable link.

Specifically we can use Bernstein's inequality \cite{boucheron2003concentration}, which involves the variance of the random samples (equal to $q(1-q)$ for our Bernoulli random success samples) and the fact that they are bounded (by $1$ in our case), to establish that 
\begin{equation}\label{eq:Bernstein}
\mathbb{P}(\hat{q}_N \leq q -\epsilon) \leq \exp\left(-\frac{N\epsilon^2/2}{q(1-q)+\epsilon/3}\right).
\end{equation}
The same bound also holds for the symmetric probability $\mathbb{P}(\hat{q}_N \geq q+ \epsilon)$. Note that this has a different behavior than Hoeffding's inequality in Lemma~\ref{lemma:Hoeffding}.

To achieve the more advantageous results we propose to employ estimation intervals for the true packet success rate $q$ that decay faster than before. Specifically consider Algorithm~\ref{alg:stability} and change to
\begin{align}\label{eq:fast_bounds}
q_{\min} &= \hat{q}_N - {\frac{\log(1/\delta)}{N}}, \\
q_{\max}& = \hat{q}_N + {\frac{\log(1/\delta)}{N}},
\end{align}
which decay as $1/N$ versus the $1/N^2$ that we considered before. Using the same arguments as in the proof of Theorem~\ref{thm:error_bound}, the probability that this new algorithm returns a correct answer equals
\begin{align}\label{eq:correct_stability_bern}
\mathbb{P}(A_N) = 1 - \mathbb{P}\left[ \hat{q}_N 
\leq q - \left[\left|q - 1 + \frac{1}{\rho(A)^2}\right| - {\frac{\log(1/\delta)}{N}}\right]_+
\right] .
\end{align}
Then using Bernstein's inequality \ref{eq:Bernstein}, we have that the probability of correct answer in \ref{eq:correct_stability_bern} is larger than $1-\delta$ if
\begin{equation}\label{eq:second_degree}
\frac{N\left(\left|q - 1 + \frac{1}{\rho(A)^2}\right| - {\frac{\log(1/\delta)}{N}}\right)^2 /2}
{q(1-q)+ \left(\left|q - 1 + \frac{1}{\rho(A)^2}\right| - {\frac{\log(1/\delta)}{N}}\right) /3} \geq \log(1/\delta) .
\end{equation}
This is equivalent to a second degree polynomial with respect to $N$ being larger than zero, and after a root analysis we 
can find that \ref{eq:second_degree} is satisfied if
\begin{equation}
	N \geq \frac{2 \log(1/\delta)(4 \left(\left|q - 1 + \frac{1}{\rho(A)^2}\right| + 3 q(1-q)\right)}
	{3 \left(q - 1 + \frac{1}{\rho(A)^2}\right)^2} .
\end{equation}
{From this expression we may discern the following regime.

\begin{cor}\label{cor:fast_sample_stability} 
	\emph{\textbf{(Sample Complexity for Low-Variance Channels)}} Consider the switched linear system \ref{eq:system} over an i.i.d. Bernoulli binary channel with an unknown success probability $q \in [0,1]$ and assume $q \neq 1 - \frac{1}{\rho(A)^2}$. Consider the stability analysis procedure developed in Algorithm \ref{alg:stability} with the substitution \ref{eq:fast_bounds} using $N$ i.i.d. channel samples drawn with success rate $q$ and some parameter $\delta \in (0,1)$. 
	If it holds that 
	\begin{equation}\label{eq:low_variance}
	q(1-q) \leq c \left|q - 1 + \frac{1}{\rho(A)^2}\right|, 
	\end{equation}
	for some constant $c>0$, then the procedure correctly determines the stability or instability of the system with probability $(1-\delta)$ if the number of samples is larger than
	\begin{equation}\label{eq:fast_sample_complexity}
	N \geq c' \frac{\log(1/\delta)}{\left|q - 1 + \frac{1}{\rho(A)^2}\right| }
	\end{equation}
	for some constant $c'>0$.
\end{cor}

Condition \ref{eq:low_variance} essentially means the channel success rate is as close to $0$ or $1$ as to the critical rate for system stability.
From \ref{eq:fast_sample_complexity} we see again that for channel rates close to the critical ones for stability we need a lot of samples. However comparing this with Corollary~\ref{cor:sample_stability} we observe an \textit{order of magnitude improvement} in this scaling law. In other words, we need an order of magnitude fewer samples to verify very reliable channels.
}

In the opposite regime where $q(1-q)$  is much larger than $\left|q - 1 + \frac{1}{\rho(A)^2}\right|$, then the sample complexity of the method in this section has the same rate as the method based on Hoeffidng's inequality in the previous section. In other words in that regime we do not gain anything from reducing the confidence intervals faster.

As a side remark, by reducing the confidence intervals fast as in this section, we are increasing the probability of wrong answer as defined in the previous section (event $W_N$). Hence in this case we cannot guarantee that for all $N\geq 1$ the probability that the algorithm provides a wrong answer is bounded by $\delta$ as before. This is a shortcoming of this approach and for safety reasons we focus primarily on algorithms where wrong answers are provably unlikely.

\subsection{Practical Algorithms}\label{sec:binomial}

In the previous sections we devised confidence intervals appropriately from concentration inequalities with the purpose of characterizing the finite sample complexity of the problem. For example we employed confidence intervals from Hoeffding's inequality, because it can be easily inverted to obtain (a lower bound on) the number of samples required to verify whether the system is stable or not. However these general concentration inequalities are distribution free and hence are conservative.

In practice, we can exploit the known distribution of the channel sample data, in particular the fact that $\gamma_k$ are i.i.d. Bernoulli random variables. Hence we can construct more practical confidence intervals and data-driven analysis algorithms. We expect such methods to be less conservative, that is, provide narrower confidence intervals with the same amount of data. This in turn translates to better efficiency of the data-driven stability and analysis algorithms, i.e.,  getting the correct answer more often and with fewer samples.

We see that $\sum_{k=0}^{N-1} \gamma_k$ has a binomial distribution with parameters $N$ and $q$. In particular we can exploit confidence intervals estimating the binomial proportion $q$ \cite{agresti1998approximate}. We detail a few methods next.

\textbf{Exact method.} An exact method, also known as Clopper-Pearson pearson, aims to invert the binomial distribution in order to obtain a confidence interval. This is for example the method used in Matlab functions "binofit", "paramci" to find binomial confidence intervals. Given a sequence of packet successes or failures, we can provide confidence intervals about the packet success rate as follows.
\begin{align}
q_{\min} &= \min\left\{\hat{q}\in [0,1]: F(\sum_{k=0}^{N-1} \gamma_k ; N, \hat{q} )\leq 1- \delta \right\},\\
q_{\max}& = \max\left\{\hat{q}\in [0,1]: F(\sum_{k=0}^{N-1} \gamma_k ; N, \hat{q} )\geq \delta \right\},
\end{align}
where $F(r;N,\hat{q})$ is the Binomial distribution with parameters $N, \hat{q}$ at the points $r=0,1,\ldots, N$. We note that this approach has again the guarantee that the probability of getting a wrong answer is by design less than $\delta$ for any sample size $N$ as in \ref{eq:wrong_error_bound}. A drawback of this method is that it is still practically conservative, in the sense that it gives wide confidence intervals.

\textbf{Approximate method.} A very common approximate method for finding binomial confidence intervals is using the normal approximation, also called the Wald interval. This method exploits the fact that by the Central Limit Theorem $\frac{ \sum_{k=0}^{N-1} (\gamma_k  -q)}{\sqrt{N q (1-q)}}$ behaves like a standard normal distribution in the limit. Hence this method proposes intervals
\begin{align}
q_{\min} &= \hat{q}_N - z(\delta) \sqrt{\frac{\hat{q}_N(1-\hat{q}_N)}{N}},\\
q_{\max}& = \hat{q}_N + z(\delta) \sqrt{\frac{\hat{q}_N(1-\hat{q}_N)}{N}},
\end{align}
where $z$ is the $1-\delta$ quantile of the standard normal distribution. This method is approximate and does not have the guarantee that the probability of a wrong answer is by design less than $\delta$ for any sample size $N$  as in \ref{eq:wrong_error_bound}. However practically it works well for sufficiently large number of samples and for probabilities $q$ away from the extreme 0 and 1 in the sense that it does not give very conservative confidence intervals.

We can hence plug in the confidence intervals produced by any of these methods in the stability analysis and control performance algorithms of Section~\ref{sec:problem}. {We will do that in Section~\ref{sec:numerical} numerically.} 

\subsection{Analysis for General Linear Systems}

The stability condition \ref{eq:stability_condition} of Proposition~\ref{prop:model_based} in reality holds for more general networked control architectures. In particular suppose we are interested in controlling a general linear plant with dynamics 
		\begin{equation}
		x_{k+1} = A x_k + B u_k + w_k,
		\end{equation}
		and we employ a controller of the form $u_k = K \hat{x_k}$ where $K$ is a standard LQR controller gain and $\hat{x}_k$ is an estimator built at the controller side based on information sent from the sensor at the other side. In particular suppose the sensor transmits state measurements $x_k$ over the unreliable link and the estimator evolves as
		\begin{equation}
		\hat{x}_{k} = \left\{ \begin{array}{ll}
		A \, \hat{x}_{k-1}, &\text{ if } \gamma_{k}=0 \\
		x_k, &\text{ if } \gamma_{k}=1
		\end{array}\right. .
		\end{equation}
		Then for Bernoulli packet dropping channels it is known that condition \ref{eq:stability_condition} is again necessary and sufficient for stability of the system\cite{Hespanha_survey}. As a result we may use exactly Algorithm~\ref{alg:stability} to examine the stability of this control system over an unknown channel from channel data samples.

Going even further, the system described in \ref{eq:system} is simple in the sense that the closed loop dynamics are ideally zero. More generally the approach can be extended to dynamics described by a switched linear time invariant model of the form
\begin{equation}\label{eq:general_system}
x_{k+1} = \left\{ \begin{array}{ll}
A_{c} \, x_{k} + w_{k}, &\text{ if } \gamma_{k}=1 \\
A_{o} \, x_{k} + w_{k}, &\text{ if } \gamma_{k}=0
\end{array}\right. .
\end{equation}
Here the system dynamics are described by the matrix $A_{c} \in \reals^{n \times n}$, where 'c' stands for closed-loop, and otherwise by $A_{o} \in \reals^{n \times n}$, where 'o' stands for open-loop. We assume that $A_{c} $ is asymptotically stable, implying that if system successfully transmits at each time step the state evolution $x_{k}, k\geq 0$ is stable. The open loop matrix $A_{o} $ may be unstable. 

Under an i.i.d. channel with success rate $q$ it is well known that the system is stable, i.e., $\limsup_{k \rightarrow \infty} \mathbb{E} x_k x_k^T < \infty$, if and only if the following condition holds
\begin{equation}\label{eq:general_stability_condition}
\rho\Big( \, q  \, A_c \otimes A_c + (1- q \, ) A_o \otimes A_o\Big) <1,
\end{equation}
where $\rho(.)$ denotes the spectral radius and $\otimes$ the Kronecker product. The result follows from the random jump linear system theory~\cite{Costa_Fragoso_book}, and Proposition~\ref{prop:model_based} is a special case. 
The method presented in Algorithm~\ref{alg:stability} can be extended to this more general case. For a fixed number of samples we can construct an interval where the true packet success rate $q$ lies with a desired high confidence level. Then we can check whether the above stability condition \ref{eq:general_stability_condition} holds for all values $q$ in the interval. If this holds, we can declare stability. Symmetrically, if for all values of the packet success $q$ within the high-confidence interval the above condition \ref{eq:general_stability_condition} does not hold, then we can declare instability. As before, as the number of samples increases the probability that this procedure provides the correct answer increases to one. The caveat is that the above stability condition \ref{eq:general_stability_condition} is in general non-convex in the variable $q \in [0,1]$, so checking whether it holds for all values $q$ in an interval may be computationally demanding. In special cases the procedure may be simplified, for example when $A_c = 0$ this boils down to the simple case we have examined in this paper. 

In this paper for simplicity of exposition we examine the simple system dynamics \ref{eq:system}. We point out that there is an extensive literature on more general model-based analysis of networked systems, e.g., when other phenomena are taken into account or different sensing and controller structures \cite{Hespanha_survey} and we believe our analysis can be extended in such scenarios.

\section{Control Performance Analysis Using Channel Samples}\label{sec:control_performance}

Beyond verifying stability we may be interested in collecting channel samples in order to verify the control performance of the system over the unknown channel, supposing the system is stable. Formally, given a finite number of channel samples we are interested in verifying with accuracy whether the control cost of the system is below some required bound $J(q) \leq J_{\text{req}}$. Similar to the previous section we can define the event that an algorithm returns the correct answer and the event that the algorithm returns a wrong answer and hence characterize the performance of the algorithm.

The proposed procedure is again based on high confidence bounds on the true channel success rate. Intuitively with the collected samples we can construct a high-confidence lower bound on the channel success $q$. Since the function $J(q)$ is {non-increasing} according to Proposition~\ref{prop:model_based}, we can check whether the control system performance under this high confidence lower bound on $q$ is lower than the desired cost value $J_{\text{req}}$. This is shown in Algorithm~\ref{alg:performance}.

\begin{algorithm}[!t]
	\caption{Performance verification}\label{alg:performance}
	\begin{algorithmic}[1]
		
		\Require Dynamics $A$, Noise covariance $W$, Desired cost level $J_{\text{req}}$, Confidence level $\delta$, Number of samples $N$, Channel samples $\gamma_0, \ldots, \gamma_{N-1} \in \{0,1\}^N$
		
		\State Compute the sample average
		\begin{equation}
		\hat{q}_N = \frac{1}{N} \sum_{k=0}^{N-1} \gamma_k
		\end{equation}
		\State Compute the high confidence lower and upper bounds
		\begin{align}
		q_{\min} &= \max\{0, \hat{q}_N - \sqrt{\frac{\log(1/\delta)}{2 N}} \}\\
		q_{\max}& = \min\{1, \hat{q}_N + \sqrt{\frac{\log(1/\delta)}{2 N}} \}
		\end{align}

		\If  {there exists a matrix $P$ such that
			\begin{subequations}\label{eq:check_feasibility}
				\begin{align}
				&\text{Tr}(PW) \leq J_{\text{req}}, \\ &P = Q + (1-q_{\min}) A^T P A, \\ &P \succeq 0.
				\end{align}
			\end{subequations}

		}
		\State\Return 'Cost is less than $J_{\text{req}}$'
		\Else
		\If {there exists a matrix $P$ such that 
			\begin{subequations}\label{eq:check_infeasibility}
				\begin{align} 
				&\text{Tr}(PW) \geq J_{\text{req}}, \\ &P = Q + (1-q_{\max}) A^T P A, \\ &P \succeq 0.
				\end{align}
			\end{subequations}
		}
		\State\Return 'Cost is larger than $J_{\text{req}}$'
		\Else
		\State\Return 'Undetermined' 
		\EndIf
		\EndIf
	\end{algorithmic}
\end{algorithm}

\begin{thm}  Consider the switched linear system \ref{eq:system} over an i.i.d. Bernoulli binary channel with an unknown success probability $q \in [0,1]$ and assume $q > 1 - \frac{1}{\rho(A)^2}$. Let $q^*$ be the optimal solution to the problem
	\begin{subequations} \label{eq:q_tilde}
		\begin{align}
		&\text{minimize} &&\tilde{q}\\
		&\text{subject to}
		&&\text{Tr}(PW) \leq J_{\text{req}},\\
		&&&P = Q + (1-\tilde{q}) A^T P A, \\ 
		&&&P \succeq 0, \, 0 \leq \tilde{q} \leq 1, 
		\end{align}
	\end{subequations}
	and assume $q \neq q^*$. Consider the control cost analysis procedure developed in Algorithm \ref{alg:performance} using $N$ i.i.d. channel samples drawn with success rate $q$ and some parameter $\delta \in (0,1)$. Then:
	\begin{enumerate}
		\item for any number of samples $N$ the procedure returns a wrong answer with probability less than $\delta$,
		\item if the number of samples is 
		\begin{equation}\label{eq:sample_complexity_for_performance}
		N \geq \frac{2 \log(1/\delta)}{ (q- q^*)^2 },
		\end{equation}
		then the procedure returns a correct answer with probability at least $(1-\delta)$.
	\end{enumerate}
\end{thm}

The proof is included next and is primarily based on similar arguments as in the previous section.

\begin{pf*}{Proof.}
We consider the case where the cost of the system is indeed lower than the desired value, i.e., $J(q) \leq J_{\text{req}}$ -- a symmetric argument verifies the case when the cost is larger that the desired value. Note by Proposition~\ref{prop:model_based} and the expression for the control cost $J(q)$ given in \ref{eq:control_cost}, we have that the value $q$ is feasible for problem \ref{eq:q_tilde}, and since $q^*$ is the optimal solution for problem \ref{eq:q_tilde}, this is implies that $q \geq q^*$, and since we have assumed that  $q \neq q^*$, we have that $q> q^*$.

\textit{Proof of (2)}: First we show that the algorithm returns the correct answer with high probability. Following the same arguments as in the proof of Theorem \ref{thm:error_bound} and Corollary~\ref{cor:sample_stability} and since $q > q^*$, if the number of samples satisfies \ref{eq:sample_complexity_for_performance}, then the event $q^* < q_{\min}$ occurs with high probability (at least $1-\delta$). Then since $q^*$ is the optimal solution to problem \ref{eq:q_tilde}, this also implies that the data-dependent random variable $q_{\min}$ is feasible for problem \ref{eq:q_tilde} with high probability. Hence there exists a matrix $P_{\min} \succeq 0$ such that $P_{\min} = Q + (1-\tilde{q}) A^T P_{\min} A$ and $\text{Tr}(P_{\min}W) \leq J_{\text{req}}$. This matrix makes also problem \ref{eq:check_feasibility} feasible, and this proves that the algorithm returns the correct answer with high probability (at least $1-\delta$).

\textit{Proof of (1)}: Second we will show that the algorithm returns the wrong answer with low probability. For that we make use of the following fact:

\textbf{Fact:} If problem \ref{eq:check_infeasibility} is feasible, then $q_{\max} \leq q^*$. 

We have also argued that $q^*< q$. So if problem \ref{eq:check_infeasibility} is feasible, then we combine the last two statements to conclude that $q_{\max} \leq q$. As a result the event that \ref{eq:check_infeasibility} is feasible implies the event that $q_{\max} \leq q$. Hence we can bound the probability that the algorithm returns the wrong answer as
\begin{equation}
	\mathbb{P}(\text{\ref{eq:check_infeasibility} feasible})\leq \mathbb{P}(q_{\max} \leq q) \leq \delta
\end{equation}
where the latter follows by Hoeffding's inequality (Lemma \ref{lemma:Hoeffding_2}).

To complete this proof we need to prove the above Fact. Note that if problem \ref{eq:check_infeasibility} is feasible, then that means that the cost of the system on a channel with rate $q_{\max}$ is larger than $J_{\text{req}}$, i.e., $J(q_{\max}) \geq J_{\text{req}}$. On the other hand by design of problem \ref{eq:q_tilde} we have that $q^* = \argmin\{\tilde{q}: J(\tilde{q})\leq J_{\text{req}}\}$. Combining the two statements we conclude that $J(q_{\max}) \geq J_{\text{req}} \geq J(q^*)$. Finally we have shown in Proposition~\ref{prop:model_based} that the function $J(q)$ is strictly decreasing. Hence $q_{\max}\leq q^*$ and this concludes the above fact.
\qedsymbol\end{pf*}

{ The problem \ref{eq:q_tilde} essentially searches for the minimum channel quality value that makes the control cost smaller than the desired bound, i.e., $q^* = \argmin\{q\in [0,1]: J(q)\leq J_{\text{req}}\}$. We note that the optimization problem \ref{eq:q_tilde} is not  jointly convex in the variables $q,P$ but can be solved with bisection. In contrast both problems \ref{eq:check_feasibility} and \ref{eq:check_infeasibility} can be solved as convex optimization problems.}

Interestingly the proposed cost performance analysis has qualitatively the same form of sample complexity as the stability analysis according to Corollary~\ref{cor:sample_stability}, even though these are two different questions. More specifically we see that what matter is the difference between the channel conditions $q$ and the solution to the problem \ref{eq:q_tilde}. Again we see that properties of the networked system over the channel adversely affects the required amount of samples.

{We note here that unlike the stability verification question of the previous section, here even if the system under consideration is asymptotically stable ($\rho(A)<1$), testing whether its performance satisfies a desired value may still require a very large number of samples.}

\section{Numerical simulations} \label{sec:numerical}

We consider a system of the form \ref{eq:system} with spectral radius $\rho(A) = 2$ that evolves over a Bernoulli channel with success rate $q = 0.9$. For this values the system is stable because \ref{eq:stability_condition} holds. For $1000$ trials we draw $N = 2000$ i.i.d. channel samples according to the success rate $q$. For each trial we run the stability test described in Algorithm~\ref{alg:stability}.

\begin{figure}[!t]
	\centering
	\includegraphics[width=1.0\columnwidth]{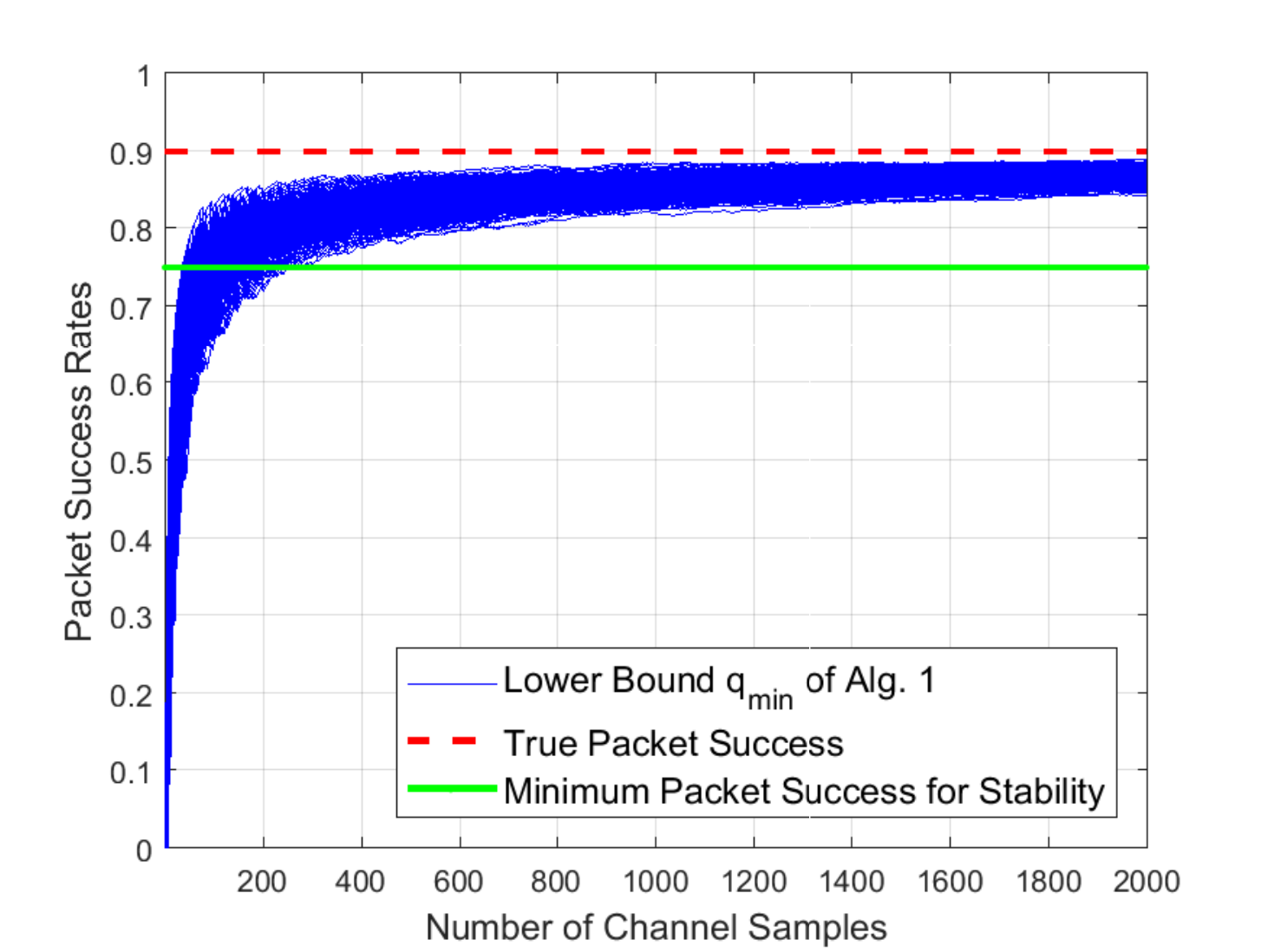}
	\vspace{-0pt}
	\caption{We consider a system and channel that lead to stability. For different trials we plot the value of the high-confidence lower bound on the true packet success rate $q$ computed by Algorithm~\ref{alg:stability} as the number of samples $N$ grow. These lower bounds grow on average above the minimum packet success rate required for stability as described in Proposition~\ref{prop:model_based}, also plotted. As the number of channel samples grows, on average the algorithm correctly verifies the stability of the system.}
	\label{fig:trials}
	\vspace{-0pt}
\end{figure}

In Fig.~\ref{fig:trials}, for different trials we plot the value of the high-confidence lower bound $q_{\min}$ on the true packet success rate $q$ computed by Algorithm~\ref{alg:stability} as the number of samples $N$ grow. These lower bounds converge to the true packet success, also plotted in the figure. We also plot the minimum packet success rate required for stability which is $1 - 1/\rho(A)^2$ as described in Proposition~\ref{prop:model_based}. Algorithm~\ref{alg:stability} checks stability by checking whether the lower bounds exceed the minimum packet success rate. As the number of channel samples grows, on average more of the lower bounds exceed the threshold and the algorithm correctly verifies the stability of the system. In Fig.~\ref{fig:trials} this takes about $300$ channel data samples.

\begin{figure}[!t]
	\centering
	\includegraphics[width=1.0\columnwidth]{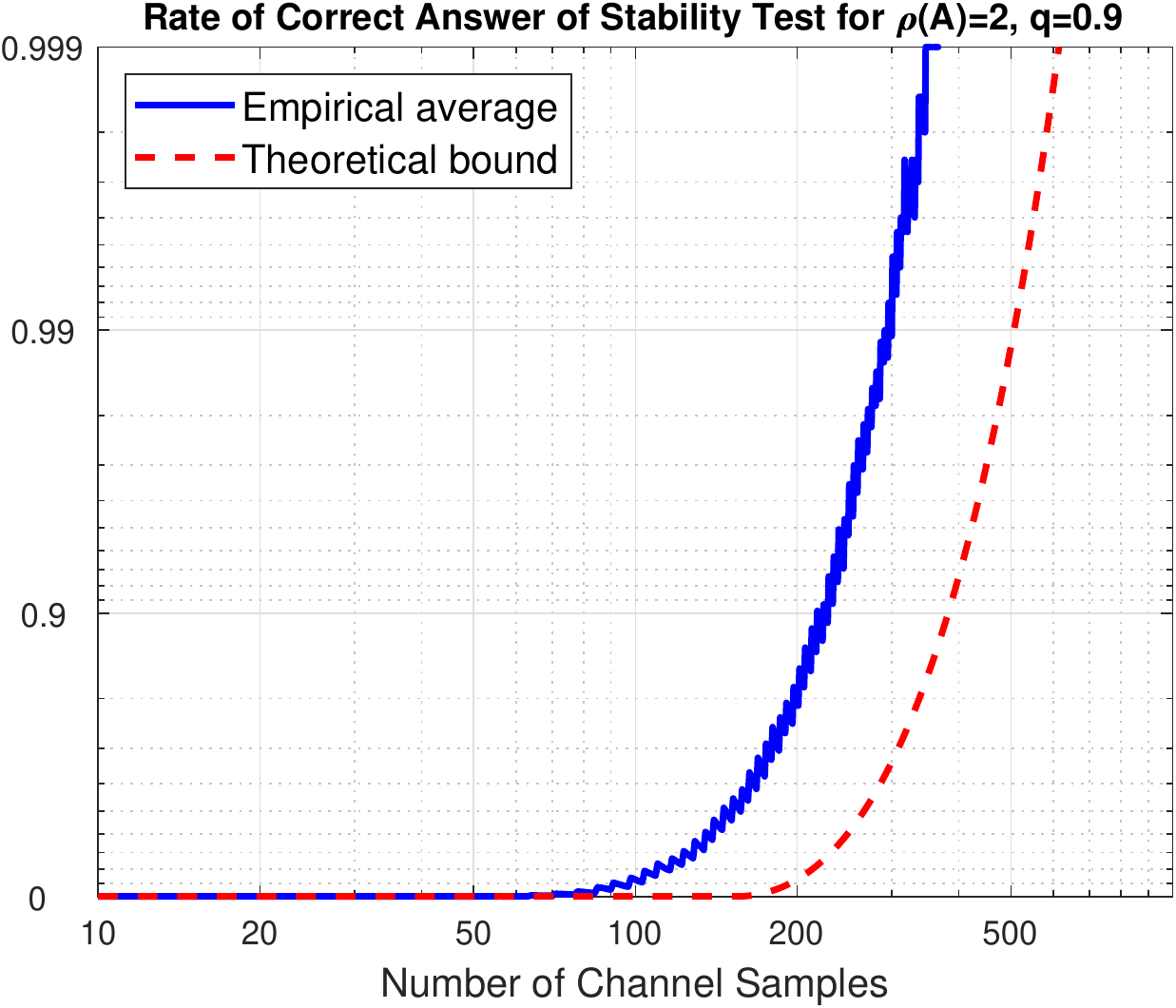}
	\vspace{-0pt}
	\caption{We consider a system and channel that lead to stability. As the number of channel samples grows, the probability that Algorithm 1 correctly verifies the stability of the system grows. The theoretical bound by Theorem 1 is below the empirical bound obtained by simulation.}
	\label{fig:theoretical_comparisons}
	\vspace{-0pt}
\end{figure}

We record the responses of the algorithm as 'Unstable', 'Stable', 'Undetermined'. Across all trials we average how many times the algorithm returns the correct answer 'Stable'. This is an empirical evaluation of the correctness of the algorithm, similar to the theoretical bound described by Theorem~\ref{thm:error_bound} (cf.~\ref{eq:error_bound}). In Fig.~\ref{fig:theoretical_comparisons} we plot both the empirical average correctness of the algorithm as well as its theoretical bound as a function of the number of channel samples drawn. First, we observe that the theoretical bound indeed is a lower bound on the average correctness of the algorithm. We note also that the bound is not tight. That means that fewer channel samples are actually required to learn whether the system is stable or not than what is predicted by our theoretical bound. The reason is that Hoeffding's inequality is not tight, as already mentioned after Lemma~\ref{lemma:Hoeffding}. Empirically however the rate at which the algorithm correctly learns the system stability as the number of samples grows seems to match the rate at which the theoretical bound grows. Hence numerically it seems that our bound captures the complexity with respect to the number of samples.
{Note here that the spikes appearing in the figure are not noise due to the random samples, they appear beacause the number of packet successes is a discrete variable, not continuous.}

\begin{figure}[!t]
	\centering
	\includegraphics[width=1.0\columnwidth]{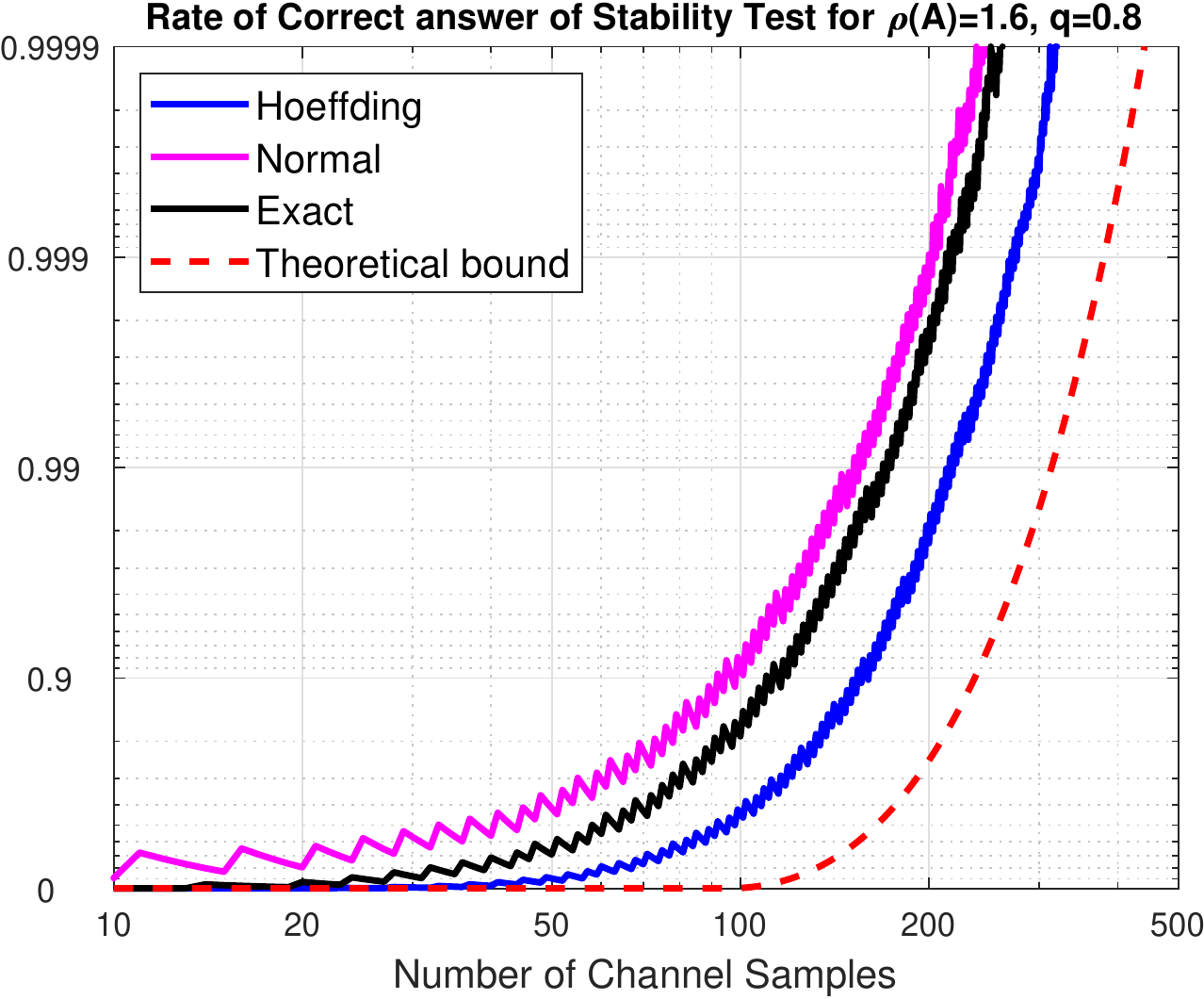}
	\vspace{-0pt}
	\caption{Comparison of stability verification algorithms using  different confidence interval methods. We compare methods based on Hoeffding;s inequality, the exact method, and the normal approximation method (Section~\ref{sec:binomial}). We empirically plot how often the stability verification algorithms returns the correct answer across a large number of iterations and number of channel samples. The normal approximation method gives the least conservative results and is more sample efficient.}
	\label{fig:comparison}
	\vspace{-0pt}
\end{figure}

{Then in Fig.~\ref{fig:comparison} we compare different algorithms for producing high confidence intervals. In particular we compare Algorithm~\ref{alg:stability} which is based on Hoeffding's inequality with the algorithms described in Section~\ref{sec:binomial} derived based on the binomial distribution. In particular we consider the algorithm based on exact confidence intervals and the algorithm based on normal approximation. We perform again an empirical evaluation of the rate of correctness of the algorithm. For multiple trials we draw $N$ i.i.d. channel samples according to the success rate $q$ and we record the responses of the algorithms as 'Unstable', 'Stable', 'Undetermined'. Across all trials we average how many times each algorithm returns the correct answer 'Stable'. We observe that the algorithm based on Hoeffding's inequality is the most conservative one. The algorithm based on exact confidence intervals has better performance, i.e., requires fewer samples on average, while the algorithm based on normal approximation provides the best performance empirically. On the other hand we observe that the average correctness of all algorithms seem to increase at similar rates with the number of samples.}

\begin{figure}[!t]
	\centering
	\includegraphics[width=1.0\columnwidth]{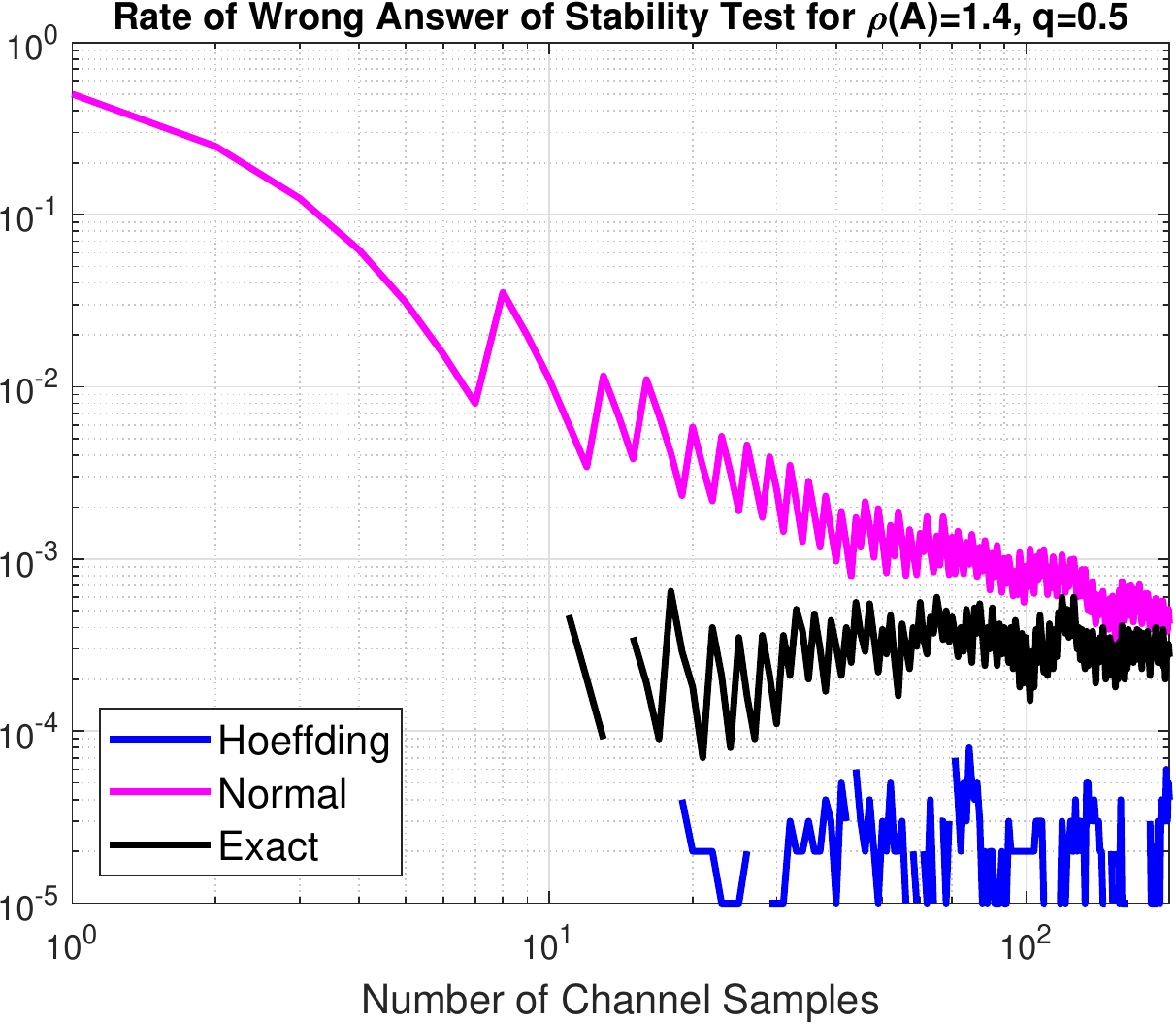}
	\vspace{-0pt}
	\caption{Comparison of stability verification algorithms using  different confidence interval methods. We compare methods based on Hoeffding's inequality, the exact method, and the normal approximation method (Section~\ref{sec:binomial}). We empirically plot how often the stability verification algorithms returns the wrong answer across a large number of iterations and number of channel samples. The normal approximation method often gives the wrong answer and especially for few number of samples.}
	\label{fig:wrong_answers}
	\vspace{-0pt}
\end{figure}

In Fig.~\ref{fig:wrong_answers} we empirically plot how often the stability verification algorithms returns the \textit{wrong answer }, i.e., return that the system is 'Unstable' while the system is stable, across a large number of iterations and number of channel samples. The normal approximation method often gives the wrong answer and especially for few number of samples. On the other hand, both the method based on Hoeffding's inequality and the exact method give the wrong answers with probability less than the desired confidence ($\delta = 10^{-3}$ here). We note that for this figure we chose a system that has very small stability margin over the channel (specifically $1-1/\rho(A)^2 \approx 0.49$ is close to $q = 0.5$ here). This shows that even in such extreme scenarios the method has guaranteed low probability of error (except for the normal approximation method for small number of samples).

\section{Discussion}\label{sec:discussion}


The assumption that the collected channel samples are i.i.d. following a Bernoulli distribution is crucial in the above results. 
In practice, only the channel sample data is available and no a priori knowledge about their distribution class, e.g., whether they are i.i.d., as in this paper, or whether they are correlated or even non-stationary. This is an important practical concern. For example, {if we utilize} the stability test developed in Algorithm~\ref{alg:stability} under i.i.d. channel assumptions on data that is not i.i.d. the probability of error in the procedure does not necessarily obey the bounds given in Theorem~\ref{thm:error_bound} above.
Ideally, we would like a more robust sample-based stability analysis. 
This is the  topic of future work.

\section{Conclusions}

In this paper we consider networked control systems over unknown channels. We utilize a learning procedure based on channel sample measurements to analyze whether the system is stable or not over a given channel, and we also analyze its control performance in terms of a quadratic control cost. As the stability analysis procedure is based on random samples there is a probability of error and we characterize this probability as a function of the true channel model, the system dynamics, and the number of collected channel samples. In particular we illustrate both theoretically and in simulations that the number of channel samples required adversely depends on the networked system stability margin, measured by the difference between the system spectral radius and the unknown channel packet success rate.

Future work involves the analysis of more complex channel models and the corresponding sample complexity, as well as controller synthesis for networked control systems over unknown channels using channel samples collected online as the system is controlled.

\bibliographystyle{ieeetr}
\bibliography{control_multiple_access,wireless_control,learning,additional}

\end{document}